\documentclass[aps,prd,superscriptaddress,floatfix]{revtex4-1}
\usepackage{color}
\usepackage{amssymb}
\usepackage[normalem]{ulem}
\usepackage{graphicx,color}
\usepackage{amsmath}
\usepackage{subfigure}
\usepackage[normalem]{ulem}
\usepackage{booktabs}
\usepackage{xcolor}
\usepackage{epsfig,graphics,color,graphicx,amsmath}

\colorlet{darkgreen}{green!50!black}
\colorlet{brightyellow}{yellow!75!red}
\colorlet{orange}{red!50!yellow}
\colorlet{darkblue}{blue!60!black}
\colorlet{darkred}{red!80!black}

\usepackage{soul}

\usepackage{mathtools}
\usepackage{mathrsfs}
\usepackage{bm}
\usepackage{relsize}
\usepackage{indentfirst}

\usepackage{xcolor}

\usepackage{amssymb,amsmath,multirow,epsfig,graphicx,color}

\usepackage{epsfig}
\usepackage{graphicx,amsmath}
\def\be{\begin{eqnarray} &&}

\def\ee{\end{eqnarray}}

\newcommand\ba{\begin{eqnarray}}
\newcommand\ea{\end{eqnarray}}

\newcommand{\bas}{\begin{eqnarray*}}
\newcommand{\eas}{\end{eqnarray*}}

\newcommand{\bno}{\begin{eqnarray*}}
\newcommand{\eno}{\end{eqnarray*}}

\def\sl
 \usepackage{hyperref}
\usepackage{url}
 \usepackage{hyperref}
\hypersetup{
	colorlinks=true,  
	linkcolor=blue,   
	filecolor=magenta,      
	urlcolor=blue,    
	citecolor=blue,   
} 
\begin{document}
\vspace{-12ex}    
\begin{flushright} 	
{\normalsize \bf \hspace{50ex}}
\end{flushright}
\vspace{11ex}
	 \title{Accelerating Universe in Terms of Hankel Function Index
	 	}
\author{K. Suratgar}
\email{suratgar.qnn@gmail.com}
\affiliation{Department of Physics, Qom Branch, Islamic Azad University, Qom, Iran}
\author{M. Mohsenzadeh}
\email{ma.mohsenzadeh@iau.ac.ir (Corresponding author)}
\affiliation{Department of Physics, Qom Branch, Islamic Azad University, Qom, Iran}
\author{E. Yusofi}
\email{eb.yusofi@iau.ac.ir}
\affiliation{Department of Physics, Ayatollah Amoli Branch, Islamic Azad University, Amol, Iran}
\affiliation{Innovation and Management Research Center, Ayatollah Amoli Branch, Islamic Azad University, Amol, Mazandaran, Iran}
\author{F. Taghizadeh-Farahmand}
\email{f$_$farahmand@qom-iau.ac.ir}
\affiliation{Department of Physics, Qom Branch, Islamic Azad University, Qom, Iran}
\date{\today}

\begin{abstract}
In this paper, $F(\nu)$ cosmology is proposed for the accelerating universe with asymptotic de Sitter  expansion in terms of Hankel function index $\nu$. To some extent, both the initial expansion during early inflation and the current accelerated expansion can be studied with a vacuum cosmic fluid i.e. $\Lambda$ in the pure de Sitter phase. Observational data further support the notion of a quasi-vacuum fluid, rather than a pure vacuum, contributing to the quasi-de Sitter acceleration in both the early and late universe.
 By examining the asymptotic expansion of the Henkel function as an approximate solution of the Mukhanov-Sasaki equation, we seek a more detailed study of quasi-de Sitter solutions in cosmology containing vacuum-like fluid.

\noindent \hspace{0.35cm} \\
\textbf{Keywords}: Cosmological Parameters; Quasi-de Sitter Modes; $Planck$ Data; Quasi-Vacuum; Cosmic Voids
\noindent \hspace{0.35cm} \\
\\ 
\textbf{PACS}: 98.80.Bp, 98.80.Jk, 98.80.Es, 98.80.Qc	
\end{abstract}

\maketitle

\section{Introduction and Motivation}
\par Inflation is the first period of rapid cosmic expansion in the very early universe. Its exponential expansion is thought to be responsible for the present large-scale homogeneity and isotropy \cite{Guth:1980zm,Liddle:1999mq}. The important achievement of this scenario is the prediction of the almost constant power spectrum of the primary quantum fluctuations. This theoretical prediction is supported by what we have measured today in cosmic fluid perturbations. With the limited information that we can extract from the analysis of these spectra, we will be faced with a wide range of acceptable inflation scenarios, but unfortunately, we have not reached a unified model of it up to now \cite{Guth:2003rn,Linde:2017pwt}.\\
Another accelerated phase of the expansion of the universe is related to the post-matter-dominant phase during the current universe \cite{Nojiri:2004pf,Nojiri:2017ncd}. "Dark energy" is an unknown component of an energy source that causes this late cosmic acceleration. Observational data from Type Ia supernovae (SNe), large-scale structure (LSS), baryon acoustic oscillations (BAO), and cosmic microwave background radiation (CMBR) have confirmed the existence of large-scale accelerative expansion with very good accuracy \cite{Amendola:2010dto}.\\
The recent CMB results from Planck satellite \cite{Planck:2013jfk,Planck:2015sxf,Planck:2018jri}, and data from the
Wilkinson Microwave Anisotropy Probe (WMAP), impose an important
constraint on the value of scalar spectral index. The most recent constrains on $n_s$ and tensor-to-scalar ratio $r$, are as follows \cite{Planck:2018jri},
$$n_s = {0.9647} \pm {0.0044}\quad , \qquad r < 0.16$$ . In
\cite{Suratgar:2022tsi}, we have considered this constraint to show that the index
of Hankel function, lies in the range of $$1.5076\leq \nu \leq 1.5098$$.\\
The standard Bunch-Davies (BD) mode \cite{Baumann:2009ds} is used just for the
pure de Sitter(dS) space-time i.e., it is obtained by exactly
setting $\nu={1.5000}$, thus the above mentioned range of index $\nu$,
motivate us to consider quasi-de Sitter modes with  $\nu\neq{1.5000}$. In our previous works \cite{Mohsenzadeh:2013bba,Mohsenzadeh:2014zfa,Yusofi:2014mma} and \cite{Yusofi:2014mta,Yusofi:2014pva,Yusofi:2018lqb,Yusofi:2018udu}, we utilized an approximate method as well as the asymptotic expansion of Hankel functions to obtain these alternative modes.\\
Hankel functions $H_{\nu}^{(1)}(x) $ and $
H_{\nu}^{(2)}(x)$ , are defined by
$$ H_{\nu}^{(1)}(x)=B_{\nu}(x)+iN_{\nu}(x)$$
\begin{equation}
	H_{\nu}^{(2)}(x)=B_{\nu}(x)-iN_{\nu}(x),
\end{equation}
where $i=\sqrt{-1}$. These two linear combinations are also known as
\emph{Bessel functions of the third kind}. $B_{\nu}(x)$ and $N_{\nu}(x)$ are the Bessel functions of the
first and second kind, respectively \cite{Arfken:1985gbs,Abramowitz:1974hat}. Depending on the positive or negative sign of the frequency, the physical application of the Hankel functions is to express outward and inward propagating cylindrical or spherical solutions of the wave equation. Also, they have been used to express the general solutions of the Mukhanov-Sasaki equation during cosmic inflation in theoretical cosmology \cite{Baumann:2009ds,Baumann:2018muz,Mukhanov:2013tua}.\\
\par By considering the asymptotic expansion of the Hankel function at the far past time limit, we can derive alternative solutions for early inflation. For asymptotic de Sitter modes with $\nu\neq{1.5000}$, we have expanded the range of possibilities beyond just the pure de Sitter space-time of the standard Bunch-Davies (BD) mode. Our research focuses on these alternative modes of the early universe, and we have derived the cosmic expansion, equation of state, and power spectrum in terms of the index $\nu$. The method adopted in this study may provide new insights into the dynamics of the early universe and have important implications for cosmology from a geometrical point of view.
\\ So in section ~\ref{Sec 2.}, we will introduce initial asymptotic de Sitter modes in terms of Hankel index $\nu$, and correspondence between them and background geometry. In section ~\ref{Sec 3.},  we introduce $F(\nu)$ function and cosmological parameters in this framework in terms of index $\nu$. The observational constraints on index $\nu$ in $F(\nu)$ cosmology will be listed in two tables in section ~\ref{Sec 4.}. Modified formula for particle creation during inflation on the based of our alternative background modes is presented in section ~\ref{Sec 5.}. Some consequences and discussions of the model about primordial gravitational waves and the expansion of cosmic voids as a quasi-vacuum fluid (gas) in quasi-de Sitter phase will be presented in the final section.
\section{Initial Asymptotic de Sitter Modes}
\label{Sec 2.}
\par In terms of gauge invariant potentials $\Phi$ and $\Psi$, the diagonal metric in the perturbed form of Newtonian gauge for the early universe can be written as follows \cite{Baumann:2018muz, Linde:2007fr, Guth:2003rn},
\begin{align} \label{eq:line-element}
	ds^{2}= a^{2}(\tau)(-(1+2\Phi){d\tau}^2+(1-2\Psi){d\textbf{x}}^2).
\end{align}
The following action shows the dynamics of quantum fluctuations of the scalar inflationary field \cite{Baumann:2018muz}
\begin{equation} \label{eq:action}
	S=\frac{1}{2}\int d^3xd\tau\left((u')^2-(\nabla
	u)^2+\frac{z''}{z}u^2\right).
\end{equation}
Considering variables $u$ and $z\equiv  \frac{a \dot{\phi}}{H}$ in the Fourier space the equation of motion for inflaton perturbations can be written as the following \textit{Mukhanov-Sasaki} equation \cite{Stewart:1993bc,Baumann:2009ds}
\begin{equation}\label{muk7}
	{u}_{k}^{\prime \prime}+\left(k^{2}-\frac{z^{\prime \prime}}{z}\right)u_{k}=0.
\end{equation}
$u_{k}$ is the Fourier mode of quantum field and the prime symbol represents the derivative with respect to conformal time $\tau$ is in the range $(-\infty, \tau_0]$. Also we can write $\frac{z^{\prime \prime}}{z}$ in terms of index $\nu$ as\cite{Stewart:1993bc, Yusofi:2014mta},
\begin{equation} \label{action2}
	\frac{z^{\prime \prime}}{z}\propto \frac{a^{\prime \prime}}{a} = \frac{\nu^{2}-1/4}{\tau^2}.
\end{equation} 
The general solution of the equation (\ref{muk7}) can be written as a linear combination of the Henkel functions of the first and second kind $H_{\nu}^{(1)}$ and $H_{\nu}^{(2)}$ as follows, \cite{Baumann:2018muz, Linde:2007fr, Guth:2003rn}:
\begin{equation}
	\label{Han22}
	u_{k}=\frac{\sqrt{\pi \tau}}{2}\Big(A_{k}H_{\nu}^{(1)}(|k\tau|)+B_{k}H_{\nu}^{(2)}(|k\tau|)\Big). \end{equation}
\subsection{Asymptotic expansion as the far past time solution}
\par Using the asymptotic expansion of the Hankel function, we can obtain an alternative solution for the equation (\ref{muk7}). The positive frequency solutions of it in terms of the Hankel index ($\nu$) and conformal time ($\tau$), can be written as follows
$$ u_{k}^{\rm ads}(\tau)= u_{k}^{\rm \nu}(\tau)= \frac{e^{-{i}k\tau}}{\sqrt{2k}}\times 
[1-i\frac{4\nu^{2}-1}{8|k\tau|}-\frac{(4\nu^{2}-1)(4\nu^{2}-9)}{2!(8|k\tau|)^2} $$
\begin{equation}
	\label{mod28} -i\frac{(4\nu^{2}-1)(4\nu^{2}-9)(4\nu^{2}-25)}{3!(8|k\tau|)^3}-...].
\end{equation}
\subsection{Introducing of F($\nu$)}
For asymptotic solutions (\ref{mod28}) up to second order of $1/k\tau$, the scale-dependent scalar power spectrum in terms of index $\nu$, and conformal time $\tau$ can obtained as following form \cite{Yusofi:2014mta},
\begin{equation} \label{del36}
	P_{s}=\frac{H^2}{(2\pi)^{2}}(\frac{H^2}{\dot{\bar{\phi}}^2})F(\nu,\tau),
\end{equation}
where we have \cite{Suratgar:2022tsi},
\begin{equation} \label{fnu}
	F(\nu,\tau)=\frac{1}{2}.\frac{2\nu+1}{(2\nu-1)}+(2\nu+1)^{2}
	\frac{(4\nu^{2}-9)^{2}}{64k^{2}\tau^{2}}+....
\end{equation}
As can be seen from relation (\ref{fnu}), for pure de Sitter phase with $\nu = \frac{3}{2}$, we obtain $F(\nu, \tau)=1$ and scale-dependency in the spectra is removed. In (\ref{fnu}), the leading term is the first one, which will have the greatest impact on the scale dependency of power spectrum at the far past time limit $k\tau \rightarrow \infty$. Therefore, we will use only this first term as $F(\nu)$ in the following form to derive cosmological parameters,
\begin{equation} \label{fnu1}
	F(\nu)=\frac{1}{2}.\frac{2\nu+1}{(2\nu-1)}.
\end{equation}
\subsection{Correspondence between background geometry and Hankel index $\nu$}
We claim that the parameter $\nu$ is mutually compatible with the type of space-time geometry. To see this claim more precisely, let us consider some half-integer values of $\nu$ in generalized solutions (\ref{mod28}) \cite{Heydarzadeh:2021uht, Yusofi:2014pva},
\begin{equation}
	\label{Bun2930} u_{k}^{\pm1/2}=\frac{1}{\sqrt{2k}}e^{-ik\tau},
\end{equation}
\begin{equation}
	\label{Bun293} u_{k}^{\pm3/2}=\frac{1}{\sqrt{2k}}(1-\frac{i}{k\tau})e^{-ik\tau},
\end{equation}
\begin{equation}
	\label{nonds1} u_{k}^{\pm5/2}=\frac{1}{\sqrt{2k}}(1-\frac{3i}{k\tau}-\frac{3}{k^2\tau^2})e^{-ik\tau},
\end{equation}
\begin{equation}
	\label{nonds2} u_{k}^{\pm7/2}=\frac{1}{\sqrt{2k}}(1-\frac{6i}{k\tau}-\frac{15}{k^2\tau^2}-\frac{15i}{k^3\tau^3})e^{-ik\tau}.
\end{equation}
\par It is clearly seen that the wave function $u_{k}^{\pm1/2}$ represents the flat space-time mode function and $u_{k}^{\pm3/2}$ represents the pure de Sitter space-time one. Other wave functions can be represented by a wave equation with a quasi-de Sitter geometry.  
\section{Parameters in F($\nu$) Cosmology}
\label{Sec 3.}
\par In this section, we will show that some important parameters in the early and late universe can be written as a function of index $\nu$. Based on the correspondence between background geometry and Hankel index $\nu$, this direct relationship can mean that all these parameters can be changed with variation of background geometry. To see the details of the calculations, the reader can refer to \cite{Suratgar:2022tsi}.\\
\subsection{Tilt}
To get the tilt, we start with the following formula\cite{Baumann:2009ds},

\begin{equation}
	n_s-1=\frac{{d\,\ln {P_s}}}{{d\,\ln k}} = \frac{{d\,\ln {P_s}}}{{d\,N}} \times \frac{{d\,N}}{{d\,\ln k}}
\end{equation}
After performing a series of straightforward and lengthy calculations (see appendix in \cite{Suratgar:2022tsi}), tilt is obtained as
\begin{equation} \label{ns1}
	n_{s}-1 = 4\left(\frac{2\nu-3}{2\nu-5}\right)-\frac{16(2\nu-3)^2}{(4\nu^2-1)(5-2\nu)^3}.
\end{equation}
As expected for the de Sitter limit ($\nu = \frac{3}{2}$), we obtain $n_s = 1$ as the scale-invariant spectra. In table \ref{table:I}, the values of $n_s$ for our model is listed for observational values of $\nu$.\\
\subsection{Slow-roll parameter}
According to Hubble's law, the speed at which galaxies move away from each other, denoted as $v$, is directly proportional to their distance, represented by $d$. This relationship can be described by the equation
\begin{equation} \label{hard909}
	v = H{d}~,
\end{equation} 
where H is known as the Hubble parameter. This parameter is given as follows in terms of the first time derivative of the scale factor,
\begin{equation} \label{hard908}
	H = \frac{\dot{a}}{a}
\end{equation}
From (\ref{action2}) and $dt = a(\tau)d\tau$, we can obtain scale factor in terms of cosmic time $t$ as \cite{Yusofi:2014pva},
\begin{equation} \label{atnu}
	a(t)\propto t^{\frac{1/2\pm\nu}{3/2\pm\nu}}.
\end{equation}
So in term of $\nu$ we obtain from (\ref{hard908}) and (\ref{atnu}),
\begin{equation} \label{hard907}
	H =\frac{1}{t}
	(\frac{1-2\nu}{3-2\nu}).
\end{equation}
Now, let's calculate the slow-roll parameter in terms of $\nu$ index. We start by defining it in terms of Hubble parameter as,
\begin{equation}	\label{eta30} 
	\epsilon = -\frac{{\dot{H}}}{H^2}.
\end{equation}
In terms of the conformal time $\tau$, the slow-roll parameter read \cite{Baumann:2018muz}
\begin{equation}
	\label{eta31} \epsilon = 1-\frac{\textbf{H}'}{\textbf{H}^2}.
\end{equation}
where $\textbf{H} = a{H}$, is the conformal Hubble parameter. For both scalar and tensor fluctuations in the power law inflation, the slow-roll parameter $\epsilon$ relates in the same form to index $\nu$  as follows \cite{Heydarzadeh:2021uht},
\begin{equation} \label{e1}
	\epsilon = \frac{2\nu-3}{2\nu-1}.
\end{equation}
\subsection{Tensor to scalar ratio}
On the other hand, for tensor perturbations (gravitational wave), we will obtain power spectrum as \cite{Suratgar:2022tsi},
\begin{equation} \label{del361}
	P_{t}=2\frac{H^2}{(2\pi)^{2}}F(\nu).
\end{equation}
Considering the $\frac{H^2}{\dot{\bar{\phi}}^2} =\frac{1}{2{\epsilon}}$  \cite{Baumann:2009ds},
and relations (\ref{del36}), and (\ref{del361}), we can obtain tensor to scalar ratio as
\begin{equation} \label{del362}
	r= \frac{P_{t}}{P_{s}}= 16\epsilon.
\end{equation}
Finally, from (\ref{e1}) we can write,  
\begin{equation} \label{del366}
	r= 16\left(\frac{2\nu-3}{2\nu-1}\right) .
\end{equation}
\subsection{Equation of state parameter}
On the other hand, in the Friedmann standard model we had a scale factor in terms of EoS parameter $w$ as follows \cite{Baumann:2018muz, Liddle:1999mq},
\begin{equation}\label{atw}
	a(t)\propto t^{\frac{2}{3(1+w)}} 
\end{equation}
By equating relations (\ref{atnu}) and (\ref{atw}), we will get the following relation between $\nu$ and $w$ as \cite{Yusofi:2014pva},
\begin{equation}
	\label{eta33} w =\frac{1}{3}\left(\frac{2\nu+3}{1-2\nu}\right).
\end{equation}
Therefore, according to (\ref{eta33}) and (\ref{Bun2930}-\ref{nonds2}), we can see that there is correspondence between values of index $\nu$, i.e. the type of background space-time geometry and the values of parameter $w$.  It seems that type of cosmic fluid may be dependent on the type of background spacetime and vice versa \cite{Yusofi:2018lqb}.\\
\subsection{Deceleration parameter}
The deceleration parameter $q$, in cosmology is defined as \cite{Linde:2014ela, Heydarzadeh:2021uht},
\begin{equation} \label{hard97}
	q=-\frac{\ddot{a}}{aH^{2}}=-\frac{a\ddot{a}}{(\dot{a})^2}.
\end{equation}
The larger value of $q$ with negative sign indicates a way to quantify the accelerated expanding at any time $t$. Therefore from (\ref{atnu}), (\ref{hard907}) and (\ref{hard97}), we can obtain deceleration parameter as following form in terms of index $\nu$,
\begin{equation} \label{hard909}
	q = \frac{2}{1-2\nu}.
\end{equation}
In table \ref{table:I}, the values of deceleration parameter $q$ is listed in terms of index $\nu$. At the beginning of inflation, the deceleration parameter equal to $-1$, but as the index $\nu$ increases, the acceleration of the universe decreases and $ q \rightarrow 0$.

In tables \ref{table:I} and \ref{table:II}, the cosmological parameters are listed for observational and half-integer values of index $\nu$. In the pure de Sitter inflation $\nu = 1.50$, the slow-roll parameter is equal to zero, but as the index $\nu > 1.50$ increases, the acceleration of the universe decreases to zero $q \rightarrow 0$ and the slow-roll parameter increase to 1. At the pure de Sitter inflation, the EoS parameter is equal to $-1$, but as the index $\nu$ increases, the EoS parameter is asymptotically close to $(\frac{-1}{3})$. The significant result from these two tables is that a quasi-de Sitter asymptotic inflation with power law expansion ($\nu > 1.50$) occurred \textit{after} pure de Sitter inflation with exponential acceleration ($\nu = 1.50$).
\section{Modified Form of Particle Creation in F($\nu$) Cosmology}
\label{Sec 5.}
Using the Bogoliubov coefficients\cite{Farley:2005px}, the standard formula ( ${\rm sta}$ ) for the initially created particles in terms of the vacuum modes $ u_{k} $ has been given by \cite{Mijic:1998if,Pereira:2009kv,RodriguezRoman:2020lra},
\begin{equation} \label{equ13} \langle N \rangle_{\rm sta}= \frac{1}{4\omega_{k}(\tau)}|{u}'_{k}(\tau)|^{2}+\frac{\omega_{k}(\tau)}{4}|{u}_{k}(\tau)|^{2}-\frac{1}{2}. 
\end{equation}
Usually, particles are created due to changes in the gravitational field caused by a curved or expanding background. In the standard view, flat space-time is considered to be the background that has the lowest energy and the minimum number of particles. Therefore, if we put the flat space-time state ($\frac{1}{\sqrt{2k}}e^{-ik\tau}$) instead of the state $u_{k}(\tau)$ in the relation (\ref{equ13}), the result $\langle N \rangle_{\rm sta}$ vanishes. This means that in the standard method, the value $\frac{1}{2}$ in the formula is due to choosing a flat background.
\begin{equation}
	\label{equ14} \langle N \rangle_{\rm flat}= \frac{1}{4\omega_{k}(\tau)}|{u_{k}^{\pm1/2}}'(\tau)|^{2}+\frac{\omega_{k}(\tau)}{4}|{u_{k}^{\pm1/2}}(\tau)|^{2}= \frac{1}{2}
\end{equation}
As a result of above discussion, we can write \cite{Ziyaee:2020wik}:
\begin{equation}\label{equ141}\langle N \rangle_{\rm sta}= \langle N \rangle_{\rm phy}-\langle N \rangle_{\rm bac}= \langle N \rangle_{\rm ads}-\langle N \rangle_{\rm flat}
\end{equation}
So in terms of physical (phy), asymptotic de Sitter (ads) and background (bac) vacuum modes $u_{k}(\tau)$, we can write the relation (\ref{equ13}) as follows
\begin{equation} \label{equ15} \langle N \rangle_{\rm sta}= \frac{1}{4\omega_{k}(\tau)}|{u^{\rm ads}}'_{k}(\tau)|^{2}+\frac{\omega_{k}(\tau)}{4}|{u^{\rm ads}}_{k}(\tau)|^{2}-\frac{1}{4\omega_{k}(\tau)}|{u_{k}^{\rm flat}}'(\tau)|^{2}+\frac{\omega_{k}(\tau)}{4}|{u_{k}^{\rm flat}}(\tau)|^{2}. 
\end{equation}
Choosing flat background in (\ref{equ141}) has two following problems;\\ (1) Covariance Problem: The formula for calculating of the number of particles in (\ref{equ141}) does not have covariance under curved space-time symmetry \cite{Ziyaee:2020wik,Ziyaee:2021pno}. \\
(2) Negative Number Problem: According to relation (\ref{equ15}), for some values of $\nu$ that related some curved space-times, we reach a negative value for the number of particles i.e. $\langle N \rangle_{\rm sta} < 0$. See this issue more clearly in \cite{Yusofi:2018udu, Ziyaee:2020wik}.\\
Our suggestion to modify of relation (\ref{equ15}) has two steps. First, both vacuum modes must be selected from curved (cur), and the covariant (cov) form of the formula changes as,
\begin{equation}\langle N \rangle_{\rm cov}= \langle N \rangle_{\rm phy}-\langle N \rangle_{\rm cur},
\end{equation}
The second step in our efforts is to solve the issue of negative number particles. It is crucial for us to consider the background space-time that is related to the one with the minimum number of particles, as highlighted in the \cite{Ziyaee:2021pno}. Based on the analysis of this paper, we have developed a final modified (mod) formula for particle creation during asymptotic de Sitter inflation. The result formula is as follows:
\begin{equation}\langle N \rangle_{\rm mod}= \langle N \rangle_{\rm ads}-\langle N \rangle_{\rm min}
\end{equation}
or in terms of vacuum modes \cite{Ziyaee:2021pno},
\begin{equation} \label{equ19} \langle N \rangle_{\rm mod}= \frac{1}{4\omega_{k}(\tau)}|{u^{\rm ads}}'_{k}(\tau)|^{2}+\frac{\omega_{k}(\tau)}{4}|{u^{\rm ads}}_{k}(\tau)|^{2}-\frac{1}{4\omega_{k}(\tau)}|{u^{\rm min}}'_{k}(\tau)|^{2}+\frac{\omega_{k}(\tau)}{4}|{u^{\rm min}}_{k}(\tau)|^{2}. 
\end{equation}
\begin{table}
	\caption{Some cosmological parameters for observational ranges of index $\nu$ based on \textit{Planck} data}
	\label{tab:1}       
	\begin{center}
		\begin{tabular}{||c||c|c|c|c|c|c|}
			\hline
			$\nu$ &\quad $1.5000$ &\quad $1.5020$&\quad $1.5040$ &\quad $1.5060$ &\quad $1.5080$ &\quad $1.5100$\\
			\hline
			\hline
			$n_s$ &\quad $1.0000$ &\quad $0.9919$&\quad $0.9839$ &\quad $0.9758$ &\quad $0.9677$ &\quad $ 0.9595$\\
			\hline
			$\epsilon$ &\quad $0.0000$ &\quad $0.0020$&\quad $0.0040$ &\quad $0.0060$ &\quad $0.0080$ &\quad $0.0100$\\
			\hline
			$r$ &\quad $0.0000$ &\quad $0.0319$&\quad $0.0637$ &\quad $0.0954$ &\quad $0.1270$ &\quad $ 0.1584$\\
			\hline
			$w$ &\quad $-1.0000$ &\quad $-0.9987$&\quad $-0.9973$ &\quad $-0.9960$ &\quad $-0.9947$ &\quad $ -0.9934$\\
			\hline
			$q$ &\quad $-1.0000$ &\quad $-0.9980$&\quad $-0.9960$ &\quad $-0.9940$ &\quad $-0.9920$ &\quad $ -0.9900$\\
			\hline

		\end{tabular}
	\end{center}
	\label{table:I}
\end{table}
\begin{table}
	\caption{Some cosmological parameters for half-integer values of index $\nu$}
	\label{tab:2}       
	\begin{center}
		\begin{tabular}{||c||c|c|c|c|c|}
			\hline
			$\nu$ &\quad $1.50$ &\quad $2.50$ &\quad $3.50$ &\quad $4.50$ &\quad $+\infty$\\
			\hline
			\hline
			$\epsilon$ &\quad $0.00$ &\quad $0.50$ &\quad $0.67$ &\quad $0.75$ &\quad $ 1.00$\\
			\hline
			$w$ &\quad $-1.00$ &\quad $-0.67$ &\quad $-0.56$ &\quad $-0.50$ &\quad  $-0.33$ \\
			\hline
			$q$ &\quad $-1.00$ &\quad $-0.50$ &\quad $-0.33$ &\quad $-0.25$ &\quad $ 0.00$\\
			\hline
			
		\end{tabular}
	\end{center}
	\label{table:II}
\end{table}

\section{Observational constraints on F($\nu$) Cocmology}
\label{Sec 4.}
\par In this section, we will use the constraints related to $n_s$ and $r$ from Planck to obtain the observational constraints on the index $\nu$.\\ $Planck$ TT,TE,EE$+$lensing$+$lowEB  from table 3. of $Planck$ 2018 \cite{Planck:2018jri}
\begin{equation} \label{nsp18}
	n_s = 0.9647 \pm 0.0044 \quad , \qquad r < 0.160
\end{equation}
The table \ref{table:I}, includes the allowable range of $\nu$. Upon comparing the data from Planck 2013, 2015, and 2018, it can be observed that the range of change for $\nu$ during the early acceleration epoch is as follows \cite{Suratgar:2022tsi}:
\begin{equation}
	\label{nuinf18} 1.5076\leq \nu_{\rm 2018} \leq 1.5098.
\end{equation}
Also, based on Planck's upper bound for $r$ (\ref{nsp18}), the maximum limit for $\nu$ are obtained as,
\begin{equation} \label{v18}
	\nu_{\rm max_{2018}} = 1.5100.
\end{equation}
For the early accelerated universe based on (\ref{eta33}), the constraints on the choice of the equation of state parameter $w$ based on constraints on Hankel function index ( \ref{nuinf18}) are given as
\begin{equation}
	\label{v39}
	-0.9959 \leq w_{\rm early} \leq -0.9895,
\end{equation}

On the other hand, for the late universe according to recent $Planck$ data, the constraints on the dark energy equation of state $w_{\rm late}$, is generally as follows \cite{Planck:2013jfk, Planck:2015sxf,Planck:2018jri},
\begin{equation}
	\label{wdark} -0.9610 \leq w_{\rm late} \leq -1.2300,
\end{equation}
\par Based on observational constraints (\ref{v39}) and (\ref{wdark}), it is suggested that the cosmic fluid originated from a quasi-vacuum phase ($w \approx -1$) during the said epoch. Furthermore, it is plausible that after billions of years, the present universe is now situated in a quasi-vacuum state once again.

\section{Discussions and Conclusions}
\label{Sec 6.}
\subsection{Quasi-de Sitter Space-time, Primordal Gravitational Wave and Particle Creation}
For several reasons, in recent years, the attention of physicists, especially cosmologists, has been drawn to de Sitter space-time. Among these reasons, apart from the maximum symmetry, is the very simple geometry of this space-time, which makes it a great experimental model for studying the physics of the universe. Also, in terms of observations, cosmic inflation is a quasi-exponential expansion that results from solving de Sitter of Friedmann's equations. 
As mentioned in the introduction, the observations show that in the inflationary period, space-time is not pure de Sitter, but in the approximation of the first order, it is pure de Sitter. So with this view and according to the background field method, The quasi (asymptotic) de Sitter solutions (\ref{nonds1}) and (\ref{nonds2}) can be obtained from the effect of gravitational wave  fluctuations $\hat{h}_{\mu\nu}$ in the primordial unperturbed background $\bar{g}_{\mu\nu}$ as two following methods \cite{Mohsenzadeh:2013bba, Mohsenzadeh:2008yv,  Ziyaee:2020wik,Ziyaee:2021pno},

\begin{equation}
	\label{bfm2}
	\hat{g}_{\rm \mu\nu}= \bar{g}^{\rm flat}_{\rm \mu\nu}+\hat{h}^{\rm (2)}_{\rm \mu\nu}  
\end{equation} 
and
\begin{equation}
	\label{bfm3}
	\hat{g}_{\rm \mu\nu}=\bar{g}^{\rm dS}_{\rm \mu\nu}+\hat{h}^{\rm (1)}_{\rm \mu\nu}  
\end{equation}                 
where $\hat{h}^{\rm (1)}_{\rm \mu\nu}$ is a first order gravitational wave acts as a linear perturbation on the pure de Sitter background $\bar{g}^{\rm dS}_{\rm \mu\nu}$, while $\hat{h}^{\rm (2)}_{\rm \mu\nu}$ acts as a second order (non-linear) perturbation on the pure flat background $\bar{g}^{\rm flat}_{\rm \mu\nu}$. In the (\ref{bfm3}), the total geometry is usually taken to be quasi-de Sitter, and the background geometry is pure de Sitter. Given that the geometry of the total universe is not pure de Sitter, but in the first approximation is pure de Sitter, therefore if we do the calculations using the standard method, in fact, we have just reached the first approximation of the geometry of the universe and not the actual geometry of the total universe. If we use the method (\ref{bfm3}), although we may not have reached the true geometry of the total universe, the result is closer to the physical world than the method (\ref{bfm2}).\\
Calculations with asymptotic solutions have shown that the effect of our proposed method (\ref{bfm3}) appears only when both physical and background vacuum sets are chosen from different curved space-times. Instead of the standard way to calculate the spectra, we have proposed that the vacuum states at the initial time are selected from the curved space-time, which is asymptotically very close to de Sitter space-time at the very early time period. The calculations carried out in the framework of asymptotic dS space-time have shown that the minimum of created particles are not related to flat space-time, but in asymptotic dS space-time the number of particles created by it is less than the flat one. In the special case of c = 2.34 (please see Fig. 3 in \cite{Ziyaee:2020wik}), we have the minimum number of particle creation. As an alternative proposal, we have chosen the background vacuum based on the minimum number of created particles in that space-time rather than flat space-time and we have modified the formula for the number of particle creation.
\subsection{Quasi-Vacuum Fluid, Cosmic Voids, Dark Matter and Dark Energy}
It has been observed from present study that the equation of state parameter $w$, is gradually approaching the state of pure vacuum ($w=-1$), although the exact value of -1 has not been reported for $w$. However, based on the observational consequences of the proposed model concerning the quasi-vacuum nature of the cosmic fluid in the early and current universe, there seems to be a need for a suitable alternative to the cosmological constant ($\Lambda$) with $w=-1$ in the $\Lambda$CDM model. We wanted to bring up this important point as it may have implications for our current understanding and future research in cosmology~\cite{Yusofi:2022hhg}.
\\
According to the cosmological simulations, the vast voids in the cosmic web appear to become larger over time, resulting in a quasi-vacuum-like emptiness. However, it is worth noting that these vast voids do not burst or disappear, leading to a state of complete emptiness like a pure vacuum ($w=-1$). Recent research has suggested that the most suitable alternative to the cosmological constant, which acts as pure vacuum energy, is the surface energy of the cosmic voids acting as a quasi-vacuum objects. This hypothesis has yielded a cosmological constant of approximately $+10^{-52} {\rm m^{-2}}$, which closely aligns with findings from Planck ~\cite{Planck:2018jri,Yusofi:2022hhg}. It is important to note that cosmic voids are not completely empty like a pure vacuum. Rather, they contain a low-density gas with a small number of isolated galaxies. Consequently, the merger and expansion of these cosmic bubbles exert additional pressure on galaxies and may serve as a source of dark matter on a local scale and dark energy on a cosmological scale~\cite{Yusofi:2019sai}. Our research team has been dedicated to studying the quasi-de Sitter phase of the universe and void-dominated cosmology for an extended period of time. Our findings suggest that the merger of clusters/voids in this cosmology acts as an accelerator in the expansion rate, resembling a quasi-de Sitter phase at the cosmic scale ~\cite{Mohammadi:2023idz}. We are excited about the implications of these findings and anticipate that further related papers presenting interesting results will be forthcoming.
\section*{Acknowledgments}
This work has been supported by the Islamic Azad University, Qom Branch, Qom, Iran.

\bibliography{MPLA_Suratgar_R1.bib} 

\end{document}